\documentstyle[eqsecnum,pra,aps]{revtex}
\begin{document}
\draft
\title{A simple quantum channel having superadditivity of channel capacity}
\author{Masahide Sasaki$^\dagger$, Kentaro Kato$^{\dagger\dagger}$, Masayuki Izutsu$^\dagger$, and Osamu Hirota$^{\dagger\dagger}$}
\address{${}^{\dagger}$Communication Research Laboratory, Ministry of Posts and Telecommunications\\
 Koganei, Tokyo 184, Japan}
\address{${}^{\dagger\dagger}$Research Center for Quantum Communications, Tamagawa University\\
 Tamagawa-gakuen, Machida, Tokyo 194, Japan}
\maketitle

\begin{abstract}
When classical information is sent through a quantum channel of nonorthogonal states, there is a possibility that transmittable classical information exceeds a channel capacity in direct use of the initial channel by extending it into multi-product channel. In this letter, it is shown that this remarkable feature of a quantum channel, so-called superadditivity, appears even in as low as the third extended coding of the simplest binary input channel.  A physical implementation of this channel is indicated based on cavity QED techniques. 
\end{abstract}

\vskip1cm

Superadditivity of the classical information channel capacity is a remarkable feature in quantum communication. Namely, it is expected that more classical information can be sent through a $n$-product channel than $n$ times the amount that can be sent through a single use of a channel. Let $\{1,\cdots,N\}$ be input alphabet with respective prior probabilities $\{\xi_1,\cdots,\xi_N\}$, and let $\{\hat s_1,\cdots,\hat s_N\}$ be corresponding input quantum states, called letter states. A decoding is described by the probability operator measure (POM) $\{\hat\pi_1,\cdots,\hat\pi_{N'}\}$ corresponding to output alphabet $\{1,\cdots,N'\}$. A quantum channel is a mapping $\{1,\cdots,N\} \mapsto  \{1,\cdots,N'\}$ where quantum noises arise in decoding process itself when at least a pair of $\{\hat s_i\}$ is non-commuting which is the case considered here. For fixed $\{\xi_i\}$ and $\{\hat\pi_i\}$, the mutual information is defined as 
\begin{equation}
I(\xi:\hat\pi)=\displaystyle\sum_i \xi_i \sum_j P(j|i) \log_2 \left[\frac{P(j|i)}{\sum_k \xi_k P(j|k)}\right],
\label{eqn:mutual_info}
\end{equation}
where $P(j|i)=\text{Tr}(\hat\pi_j \hat s_i)$ is a conditional probability that the alphabet $j$ is chosen when the alphabet $i$ is 
true. The classical information channel capacity is defined as the maximum value of this mutual information obtained by optimizing $\{\xi_i\}$ and $\{\hat\pi_i\}$, 
\begin{equation}
C_1\equiv \max_{\{\xi_i\},\{\hat\pi_j\}}I(\xi:\hat\pi). 
\label{eqn:c1_def}
\end{equation}

In classical information theory, faithful signal transmission is possible by using a certain channel coding if a transmission rate $R={1\over n}\log_2 M$, where $M$ is a number of codewords and $n$ is a length of the codewords, is kept below $C_1$. In contrast, if the quantum noise in the channel is handled properly relying on quantum information theory, the transmission rate $R$ can be raised up to the von Neumann entropy, $H(\hat\rho)$,
\begin{equation}
H(\hat\rho)=-{\rm Tr}(\hat\rho\log_2\hat\rho), \quad \mbox{where  } \hat\rho=\sum_i \xi_i\hat s_i.
\label{eqn:entropy}
\end{equation}
So the von Neumann entropy is indeed the quantum channel capacity \cite{Holevo_QuantCap1979}.  This fact has recently been proved by Hausladen et. al. for a pure-state case \cite{Hausladen_Theorem}, and has been completed by Holevo including a mixed-state case \cite{Holevo_Theorem}. It is called the quantum channel coding (QCC) theorem. 

A basic channel coding consists of a concatination of the letter states in a length $n$ and a pruning of all the possible $N^n$ sequences $\{\hat s_{i_1} \otimes \cdots \otimes\hat s_{i_N}\}$ into $M$ codewords $\{ \hat S_m\vert m=1,\cdots,M\}$. Assigning an input distribution $\{\zeta_m\}$ to the codewords, the classical capacity for the above kind of $n$-th extended quantum channel can be defined as 
\begin{equation}
C_n\equiv \max_{\{\zeta_m\},\{\hat S_m\}\{\hat\Pi_j\}}I(\zeta,\hat S:\hat\Pi),  
\label{eqn:c1}
\end{equation}
where $\{\hat\Pi_j\}$ is the POM for decoding the codewords. Then, the QCC theorem means $C_n\ge nC_1$, the superadditivity of the classical capacity. 
By contrast, in classical information theory, the capacity is additive, i.e., $C_n=nC_1$.

It may be plausible that its origin is a quantum correlation among letter states, i.e., an entanglement, generated by a quantum measurement in decoding.  However, there has been no guiding principle for utilizing the entanglement correlation so as to produce the superadditivity. Even its direct and unambiguous example, not like an asymptotic one in the length $n\rightarrow\infty$, has not been found yet. 
A related work was done by Peres and Wootters \cite{PeresWootters}.  They considered three linearly dependent spin-$1\over2$ states 
$\{\vert\phi_1\rangle,\vert\phi_2\rangle,\vert\phi_3\rangle\}$ with equal prior probabilities, and studied the amount of the mutual information obtained by several kinds of quantum measurements. They showed that the mutual information obtained by using three 2-bit states 
$\{\vert\phi_1\rangle\otimes\vert\phi_1\rangle,\vert\phi_2\rangle\otimes\vert\phi_2\rangle,\vert\phi_3\rangle\otimes\vert\phi_3\rangle\}$ and by applying the combined measurement can be larger than twice of the optimum amount attained by using the three initial 1-bit letter states. This must be a gain due to a channel coding. In order to show the superadditivity, however, one must know the $C_1$ which was not given in their work. In the above kinds of linearly dependent letter states, the $C_1$ is obtained by setting one of the porior probabilities zero and the others equal, and by applying a standard von Neumann measurement, which is indeed binary quantum channel \cite{Osaki_Peres}. Then a comparison between the mutual information of an extended channel and the $C_1$ does not seem to make sense because this logic leads to a situation that all kinds of sets of more than three linearly dependent letter states may be compared with the $C_1$ of the binary channel.

An example shown in this letter would be unambiguous and more surprising. This is the simplest  case of binary input letter states, $\{ \vert+\rangle , \vert-\rangle\}$, for which identification of the classical capacity $C_1$ is established \cite{FuchsPeres,Ban_C1,Osaki_C1}. In this case, the optimization can be achieved by the binary symmetric channel with the decoding by 
$\{\hat\pi_i=\vert\omega_i\rangle\langle\omega_i\vert\}$ where
$$
\begin{array}{lcl}
\vert\omega_1\rangle&=&
       \Bigl (   {\sqrt {{1+c}\over2}} 
                   + \kappa{\sqrt {{1-c}\over{2(1-\kappa^2)}} } 
         \Bigr) \vert+\rangle
      - {\sqrt {{1-c}\over{2(1-\kappa^2)}} }\vert-\rangle, \\
\vert\omega_2\rangle&=&
        {  \sqrt {{1+c}\over{2(1-\kappa^2)}} }\vert-\rangle
        + \Bigl( {\sqrt {{1-c}\over2}} - \kappa{\sqrt {{1+c}\over{2(1-\kappa^2)}} } 
          \Bigr)\vert+\rangle, 
\end{array}
$$
\noindent
with $\kappa=\langle+\vert-\rangle$, being assumed to be real,  and $c=\sqrt{1-\kappa^2}$. Then the capacity $C_1$ is given as 
\begin{equation}
C_1=1+(1-p)\log_2 (1-p) + p\log_2 p,
\label{eqn:c1_binary}
\end{equation}
where $p=(1-\sqrt{1-\kappa^2})/2$.

Now we would like to show that the superadditivity of the classical information channel capacity reveals itself in the third-extended coding. The four sequences $\{ \vert S_i\rangle \} = \{ \vert+++\rangle,\vert+--\rangle,\vert--+\rangle,\vert-+-\rangle  \} $ are picked up as the codewords from 8 possible sequences. They can encode 2-bit classical information. We fix here their prior probabilities as $1/4$. In decoding, the so-called square-root measurement \cite{Hausladen_Theorem,Helstrom_QDET,Holevo_SubOptMeas} is applied. Let $\{\vert\mu_i\rangle\}$ be the measurement states. Giving the Gram matrix $\hat\Gamma=(\langle S_i\vert S_j\rangle)$, the channel matrix elements are then given as $x_{ij}\equiv\langle\mu_i\vert S_j\rangle=(\hat\Gamma^{1\over2})_{ij}$. It is straightforward that 
\begin{equation}
\begin{array}{lr}
x_{ii}={1\over4}(\sqrt{1+3\kappa^2} + 3 \sqrt{1-\kappa^2}),& \forall i, \\
x_{ij}={1\over4}(\sqrt{1+3\kappa^2} - \sqrt{1-\kappa^2}),& i\ne j. 
\end{array}
\end{equation}
Moreover one can confirm that this measurement attains the minimum avarage error probability, that is, $x_{ij}$ satisfies the Holevo condition \cite{Helstrom_QDET,Holevo}. The mutual information is simply given as 
\begin{equation}
I_3(S:\mu)=2+x_{11}^2\log_2x_{11}^2+3x_{12}^2\log_2x_{12}^2. 
\end{equation}
Then it can be seen that $I_3(S:\mu)/3> C_1$ for $0.74<\kappa<1$, as shown in Fig. 1 (a) and (b). This ensures the superadditivity $C_3>3C_1$.  Fig. 1 (c) shows the minimum average error probability $P_e^{(3)}({\rm opt})$. In almost the same region of $\kappa$ in which $C_3>3C_1$ holds, $P_e^{(3)}({\rm opt})$ becomes larger than the minimum average error probability, $p$, of the initial channel. Thus, while the reliability in terms of the average error rate degrades by the coding, the transmittable classical information can be raised up. 
This is in sharp contrast to a result from classical information theory that the third extention falls short of correcting even one bit error so that the obtained mutual information is far below $C_1$. 
The other combination  $\{ \vert+++\rangle,\vert-++\rangle,\vert+--\rangle,\vert---\rangle  \} $ does not show the superadditivity (see one-dotted line in Fig. 1). In the second-extended coding, the superadditivity never appears. 

Let us consider $n$-th extension. There are totally 2$^{n-1}$ sequences whose minimum Hamming distance is 2. Suppose all of them are used as codewords with equal input probabilities. Then in similar way to the above, we can calculate an accessible mutual information $I_n(S:\mu)$ by applying the square root measurement giving the Gram matrix. We have confirmed that the region of $\kappa$ where the superadditivity appears extends from $\kappa=1$ to lower value as $n$ increases. The numerical results for $n=5\sim13$ are shown in Fig. 2. It is also worth mentioning that if all of the sequences, totally $2^n$, are used as the codewords, 
$\{  \hat s_{i_1}\otimes\cdots\otimes\hat s_{i_n}  \}$ with the prior probabilities $\{  \xi_{i_1}\times\cdots\times\xi_{i_n}  \}$, the optimum decoding is  realized by 
\begin{equation}
\hat{\mit\Pi}_{i_1\cdots i_n}=\hat\pi_{i_1}\otimes\cdots\otimes\hat\pi_{i_n}
\end{equation}
for both the average error probability and the mutual information \cite{SasakiHolevo}, whose proof will be given elsewhere. In this case, the decoding process generates no entanglement among the letter states, and the resulting capacity is merely additive. Once the sequences are pruned, a decoding process may include some entanglement correlations. But necessary and sufficient conditions for inducing the superadditivity have not been clear yet.

Now let us move to a realization problem of the above kind of quantum channel, especially, an implementation of the decoding process.  So far there has been no explicite physical model corresponding to the quantum optimum decoding of codewords. We model the source $\{ \vert+\rangle , \vert-\rangle\}$ by superposition states between upper-($\vert\uparrow\rangle$) and lower-level ($\vert\downarrow\rangle$) states of a two-level atom. Namely, a series of atoms  is prepared only in $\vert\uparrow\rangle$-state, and then it passes through an encoder by which some of the atoms are transferred into $\vert\swarrow\rangle=\hat R_y(\phi)\vert\uparrow\rangle$ by a rotator 
\begin{equation}
\hat R_y(\phi)=\left(
\begin{array}{cc}
{\rm cos}{\phi\over2}   & {\rm sin}{\phi\over2}    \cr
-{\rm sin}{\phi\over2} & {\rm cos}{\phi\over2} 
\end{array}
\right).
\label{eqn:R_y}
\end{equation}
$\{ \vert\uparrow\rangle , \vert\swarrow\rangle \}$ are regarded as the letter states $\{\vert+\rangle,\vert-\rangle\}$.

We consider $n$-th extention and let ${\cal H}_{2^n}$ be the $n$-th extended Hilbert space which is spanned by an orthonormal basis states: 
\begin{equation}
\begin{array}{ccl}
\vert\uparrow\rangle\vert\uparrow\rangle\cdots\vert\uparrow\rangle\vert\uparrow\rangle&\equiv&\vert A_1\rangle, \\
\vert\uparrow\rangle\vert\uparrow\rangle\cdots\vert\uparrow\rangle\vert\downarrow\rangle&\equiv&\vert A_2\rangle, \\
      									     & \vdots & \\
\vert\downarrow\rangle\vert\downarrow\rangle\cdots\vert\downarrow\rangle\vert\uparrow\rangle&\equiv&\vert A_{2^n-1}\rangle, \\
\vert\downarrow\rangle\vert\downarrow\rangle\cdots\vert\downarrow\rangle\vert\downarrow\rangle&\equiv&\vert A_{2^n}\rangle.
\end{array}
\label{eqn:allbasis}
\end{equation}
Let $\{  \vert S_1\rangle, \cdots,   \vert S_M\rangle   \}$ $(M<2^n)$ be the codewords actually used in the channel and $\{  \vert S_{M+1}\rangle, \cdots,   \vert S_{2^n}\rangle   \}$ be the rest of them. The former set spanns the $M$-dim signal space ${\cal H}_s$. Our concern is an implementation of the square-root measurement described by $\{\vert\mu_m\rangle\vert m=1, \cdots, M\}$. $\{\vert S_i\rangle\}$ can be expanded by $\{\vert A_i\rangle\}$ as, 
\begin{equation}
\left(\begin{array}{c} \vert S_1\rangle \\
                                   \vdots \\
                                   \vert S_{2^n}\rangle    \end{array} \right)
=\hat C  \left(\begin{array}{c} \vert A_1\rangle \\
                                   \vdots \\
                                   \vert A_{2^n}\rangle    \end{array} \right),    \quad \hat C=(\langle\rho_i\vert A_j\rangle).
\end{equation}
Since $M$ codewords are linearly independent, $\{\vert\mu_m\rangle\}$ forms a complete orthonormal set on ${\cal H}_s$. Based on this set, the following orthonormal states can be introduced, 
\begin{equation}
\vert\mu_i\rangle=
{
{\vert S_i\rangle-\sum_{k=1}^{i-1}\vert\mu_k\rangle\langle\mu_k\vert S_i\rangle}
\over
{\sqrt{1-\sum_{k=1}^{i-1}\vert\langle\mu_k\vert S_i\rangle\vert^2}}  
 },  
\label{eqn:orthogonal_set}
\end{equation}
where $i=M+1,\cdots, 2^n $. We denote another expansion by $\{\vert\mu_i\rangle\vert i=1, \cdots, 2^n\}$ as, 
\begin{equation}
\left(\begin{array}{c} \vert S_1\rangle \\
                                   \vdots \\
                                   \vert S_{2^n}\rangle    \end{array} \right)
=
\hat B
\left(\begin{array}{c} \vert \mu_1\rangle \\
                                   \vdots \\
                                   \vert \mu_{2^n}\rangle    \end{array} \right).
\end{equation}
The two basis sets are connected via a unitary operator $\hat V$ as,
\begin{mathletters}
\begin{equation}
\vert S_i\rangle =\hat V^\dagger \vert A_i\rangle, \quad (i=1,\cdots,2^n),
\end{equation}
where 
\begin{equation}
\hat V^\dagger = \sum_{i,j}^{2^n} v_{ji} \vert A_j\rangle\langle A_i \vert, \quad v_{ji}=(\hat B^{-1} \hat C)_{ij}. 
\end{equation}
\label{eqn:eta_A}
\end{mathletters}

\noindent
The minimum error probability is obtained as 
\begin{equation}
P_e({\rm opt})=1-\sum_{m=1}^M \zeta_m \vert\langle\rho_m\vert \hat V^\dagger\vert A_m\rangle\vert^2, 
\end{equation}
where $\{\zeta_m\}$  is a priori distribution of the codewords. This means that the decoding by $\{\vert\mu_m\rangle\}$ can be equivalently achieved first by transforming the codewords $\{\vert S_m\rangle\}$ by the unitary transformation $\hat V $ and then by performing a von Neumann measurement $\{ \vert A_m\rangle\langle A_m\vert\}$ \cite{Sasaki's}, which is merely a level detection of individual particles (letter states).  In this scheme, what brings the entanglement among the letter states is the unitary transformation $\hat V $.

The problem is then an  implementation of $\hat V$ on the whole space ${\cal H}_{2^n}$. 
Barenco. et. al. \cite{Barenco} showed that an exact {\it simulation} of any discrete unitary operator can be carried out by using a quantum computing network. What we require here is not a {\it simulation} but rather a real {\it operation} acting on the atomic states constituting the codewords. This can  be accomplished by applying a 2-bit gate which works with {\it target} and {\it control} bits as a single atomic spieces. Sleator and Weinfurter have already proposed such a model based on the cavity QED method \cite{Sleator} (the S-W model, henceforth).

At first, $\hat V$ is decomposed into  $U(2)$-operators $\hat T_{j,i}$ \cite{Reck} as, 
\begin{mathletters}
\begin{equation}
\hat V =\hat D \hat T_{2,1}  \hat T_{3,1} \cdots \hat T_{2^n,2^n-2} \hat T_{2^n,2^n-1}, \label{eqn:decomp_a} 
\end{equation}
where 
\begin{equation}
\hat T_{j,i} ={\rm exp}[-\gamma _{ji}(\vert A_i\rangle\langle A_j\vert - \vert A_j\rangle\langle A_i\vert)]       . \label{eqn:decomp_b}
\end{equation}
\end{mathletters}

\noindent
($\langle\uparrow\vert\swarrow\rangle$ is assumed to be real.) Then the above 2-dim rotations are converted into networks of 2-bit gates by using the formula established by Barenco et. al. \cite{Barenco}. We are especially concerned with the case of $n=3$ in which the superadditivity can appear. For this case, the principle of the formula can easily be understood by showing an example, say, a rotation exp$[-\gamma(\vert\uparrow\downarrow\uparrow\rangle\langle\downarrow\uparrow\downarrow\vert - \vert\downarrow\uparrow\downarrow\rangle\langle\uparrow\downarrow\uparrow\vert)]$. It can be executed by the following network,

\vskip1cm
Diagram 1. 
\vskip1cm

\noindent
All the notations are borrowed from ref. \cite{Barenco}. 
The block denoted as $\hat M$ is for mapping $\{ \vert\uparrow\downarrow\uparrow\rangle , \vert\downarrow\uparrow\downarrow\rangle \}$ into 
$\{ \vert\downarrow\downarrow\uparrow\rangle , \vert\downarrow\downarrow\downarrow\rangle \}$.  In the mapped plane, the desired rotation is carried out as the 3-bit gate operation  
$\bigwedge_2(\hat R_y(2\gamma))$. The two 3-bit gates in the above diagram can be further decomposed into networks consisting of the 1-bit gates, 
$\bigwedge_0(\hat R_y(\pm\gamma))$ and $\bigwedge_0({\sigma_x})$, and the 2-bit gate $\bigwedge_1(\sqrt{\sigma_x})$ \cite{Barenco}. 
Implementations of the 1-bit gates are straghtforward by using the Ramsey zone (RZ) described by the following unitary operator: 
\begin{equation}
\hat U_R (\tau,\vert\epsilon\vert)=
\left(
\begin{array}{cc}
{\rm e}^{-i\nu\tau/2}{\rm cos}(\vert\epsilon\vert\tau) & {\rm e}^{-i\nu\tau/2}{\rm sin}(\vert\epsilon\vert\tau)      \cr
 -{\rm e}^{i\nu\tau/2}{\rm sin}(\vert\epsilon\vert\tau) & {\rm e}^{i\nu\tau/2}{\rm cos}(\vert\epsilon\vert\tau)
\end{array}
\right),
\end{equation}
where $\epsilon$ is a complex amplitude of a pumping field, the angular freqency $\nu$ corresponds to an atomic level separation, and $\tau$ is an interaction period.

The required 2-bit gate can be effected by  the S-W model which is modeled by the Jaynes-Cummings Hamiltonian, 
\begin{equation}
\begin{array}{lll}
\hat H=\hbar \omega {\hat a}^\dagger {\hat a} 
&+ &{1\over 2} \hbar\nu ( \vert\uparrow\rangle\langle\uparrow\vert - \vert\downarrow\rangle\langle\downarrow\vert )  \\
&+ & \hbar g ( {\hat a}^\dagger\vert\downarrow\rangle\langle\uparrow\vert 
                    + {\hat a}\vert\uparrow\rangle\langle\downarrow\vert ),
\end{array}
\end{equation}
where ${\hat a}$ (${\hat a}^\dagger$) is an annihilation (creation) operator for a cavity field with an angular frequency $\omega$, $g$ is a coupling constant between the cavity  field and the atom.
It is assumed that $\nu$ is originally detuned from the cavity resonant frequency $\omega$ so that the atom undergoes an off-resonant interaction whose time evolution is given in the spinor representation as, 
\begin{equation}
\hat U_{\rm off}(t)=\sum_{n=0}^\infty \vert n\rangle\langle n\vert
\left(
\begin{array}{cc}
{\rm e}^{-i({\nu\over2}+g_{\rm eff})t-in g_{\rm eff}t} & 0     \cr
 0 & {\rm e}^{{{i\nu t}\over2}+in g_{\rm eff}t}
\end{array}
\right),
\end{equation}
where $g_{\rm eff}=g^2/\delta$, $\delta=\nu-\omega$, and $\vert n\rangle$ is $n$-photon state. Phase factors involving $\omega$ have been  omitted since it will give no physical effect. If $\nu$ is tuned to $\omega$ by an appropriate Stark shifting, an on-resonant interaction can be carried out as, 
\begin{equation}
\hat U_{\rm on}=
\left(
\begin{array}{cc}
0 & -i  \vert 0\rangle\langle 1\vert    \cr
-i  \vert 1\rangle\langle 0\vert &  \vert 0\rangle\langle 0\vert
\end{array}
\right),
\end{equation}
where the interaction period $t_0$ is chosen as $gt_0={\pi\over2}$ and the fact is taken into account that the cavity field is either $\vert 0\rangle$ or $\vert 1\rangle$ thoughout the gate operation.  Denoting the control-, target-bit atoms and the cavity as ``a" ,``b" and ``c", respectively, $\bigwedge_1(\sqrt{\sigma_x})$ can be realized by applying a unitary process, 
$$
\begin{array}{rcl}
\hat R_z^{(a)}(-{5\over4}\pi&)&\hat R_x^{(a)}(\pi)\hat U_{\rm on}^{(a,c)} \hat U_R^{(b)} (\tau',\vert\epsilon'\vert)  \\
&\cdot&  \hat U_{\rm off}^{(b,c)}(t) \hat U_R^{(b)} (\tau,\vert\epsilon\vert)
\hat U_{\rm on}^{(a,c)} \hat R_x^{(a)}(\pi)
\label{eqn:uuu}
\end{array}
$$
where the superscript indicates on what system(s) the operator acts. Here $\vert\epsilon\vert\tau=\vert\epsilon'\vert\tau'={\pi\over4}$  and 
$$
i{\nu(\tau-\tau')\over2}-i{\nu t\over2} - i {g_{\rm eff}t\over2} = 2\pi n \quad (n={\rm integer}), 
$$
should be satisfied.

In summary, we have proposed a physical model of a quantum channel showing the superadditivity of the classical information channel capacity. It consists of four 3-bit codewords as input signals and the quantum optimum detection which can be realized as a quantum gate network based on cavity QED technique.

The authors would like to thank  Prof. A. S. Holevo of Steklov Mathematical Institute, Dr. M. Ban of Hitachi Advanced Research Laboratory, Dr. K. Yamazaki and Dr. M. Osaki of Tamagawa University, Tokyo, for their helpful discussions.

\begin{figure}
\caption{(a) The mutual information and the $C_1$  as a function of $\kappa$. The solid line represents the mutual information per bit of the channel consisting of input codewords $\{ \vert+++\rangle,\vert+--\rangle,\vert--+\rangle,\vert-+-\rangle  \}$ with equal prior probabilities and the square-root measurement for them. The one-dotted line corresponds to the case of the other input codewords $\{ \vert+++\rangle,\vert-++\rangle,\vert+--\rangle,\vert---\rangle  \}$.  
The $C_1$ (dashed line) is attained by the binary symmetric channel explaned in the text. 
(b) Same as (a), but for the region $0.7<\kappa<1$. 
(c) The minimum average error probabilities corresponding to the three kinds of channels in (a), as a function of $\kappa$. }
\label{fig1}
\end{figure}

\begin{figure}
\caption{The difference between mutual information per symble and the $C_1$  as a function of $\kappa$ for $n=5\sim13$. The region of $\kappa$ where the superadditivity appears becomes wider. }
\label{fig2}
\end{figure}

\end{document}